\documentclass[12pt]{article}
\usepackage{graphicx}
\usepackage{slashed}

\newcommand\pubnumber{SNSN-323-63}
\newcommand\pubdate{\today}

\def\met{$\slashed{E}_T$~}
\def\napoli{Physik Institut\\
Universit\"{a}t Z\"{u}rich, Switzerland}

\def\Title#1{\begin{center} {\Large #1 } \end{center}}
\def\Author#1{\begin{center}{ \sc #1} \end{center}}
\def\Address#1{\begin{center}{ \it #1} \end{center}}

\newcommand\pubblock{\rightline{\begin{tabular}{l} \pubnumber\\
         \pubdate  \end{tabular}}}
\newenvironment{Abstract}{\begin{quotation}  }{\end{quotation}}
\newenvironment{Presented}{\begin{quotation} \begin{center} 
             PRESENTED AT\end{center}\bigskip 
      \begin{center}\begin{large}}{\end{large}\end{center} \end{quotation}}

\def\beq{\begin{equation}}
\def\eeq#1{\label{#1}\end{equation}}
\def\eeqn{\end{equation}}

\def\beqa{\begin{eqnarray}}
\def\eeqa#1{\label{#1}\end{eqnarray}}
\def\eeqan{\end{eqnarray}}

\let\bar=\overbar

\def\Dslash{\not{\hbox{\kern-4pt $D$}}}
\def\dslash{\not{\hbox{\kern-2pt $\del$}}}

\def\msb{{\bar{\ssstyle M \kern -1pt S}}}

\begin{document}
\begin{titlepage}
\pubblock

\vfill
\Title{LHC results for dark matter from ATLAS and CMS}
\vfill
\Author{Annapaola de Cosa on behalf of the CMS and ATLAS collaborations}
\Address{\napoli}
\vfill
\begin{Abstract}
The CMS and ATLAS collaborations searched for Dark Matter (DM) particles directly produced in pair. The searches are performed using the full LHC Run-I dataset recorded with the CMS and ATLAS detectors in proton-proton collisions at a center-of-mass energy of 8~TeV. Signatures considered include those yielding energetic jets and large missing transverse momentum as well as electroweak bosons and heavy flavour quarks plus missing transverse energy. No deviation from SM background expectation was found and exclusion limits on DM production cross section were set.
\end{Abstract}
\vfill
\begin{Presented}
CIPANP2015: The 12th Conference on the Intersections of Particle and Nuclear Physics\\
Vail, CO (United States),  May 19--24, 2015
\end{Presented}
\vfill
\end{titlepage}
\def\thefootnote{\fnsymbol{footnote}}
\setcounter{footnote}{0}

\section{Introduction}

There is clear evidence for abundance of matter in the Universe 
that cannot be explained by the visible matter only. 
This excess of matter is called Dark Matter (DM) and constitutes 
about 25\% of the content of the Universe, while the ordinary atomic 
matter accounts only for at most 5\%. Up so far proofs of evidence 
for DM come only from the observations of the effects of its 
gravitational interaction with ordinary matter. The nature of this unknown kind of matter 
is still elusive, despite the many pieces of evidence.


Several theoretical models have been proposed, attempting 
to describe the nature of DM and its interaction with ordinary matter~\cite{Bertone}\cite{Bauer}. 
One of the most promising candidates proposed by these models are  the Weakly Interactive Massive Particles, WIMPs.


The hunt for the WIMPs involves a variety of experiments such as 
direct and indirect detection (DD and ID) experiments and searches at particle colliders. 
Direct detection of DM-SM interaction searches look for the small 
effect of DM scattering  off the target nucleus in underground experiments.
Indirect Detection experiments look for an abundance of SM particles
that may be produced from the annihilation of DM particles.
Particle collider experiments, such as Large Hadron Collider (LHC) and Tevatron experiments,
search for DM pair production in SM particle collisions.
All these search strategies are complementary as they probe different kinds of WIMP-SM interactions 
with different detectors and sensitivity.



Like neutrinos, if produced, DM would escape the detector without 
leaving signs of its passage. 
Its production can be inferred by measuring the amount of energy 
imbalance in the plane transverse to colliding beams.
The presence of further objects recoiling against WIMPs in the final state 
can be used to flag the interaction and identify it. 
A variety of signatures are exploited at LHC in order to look for DM.

In this paper results of searches for direct production of WIMPs 
with the ATLAS~\cite{ATLAS} and CMS~\cite{CMS} experiments at the LHC at CERN are presented.
The searches have been performed using $\sim$~20~fb$^{-1}$ of proton-proton collision data 
delivered during LHC Run-1 at 8~TeV centre-of-mass energy.

\section{Effective Operators }

An effective Field Theory (EFT) approach was used to interpret results from LHC Run-1 analyses: the WIMP-SM interaction is described in terms 
of contact interaction. These effective operators are characterised by the kind of interaction
(e.g. scalar, pseudo scalar, vector, axial-vector) and by two parameters of the theory:
the mass of the WIMP, m$_\chi$ and the energy scale, M$^*=M/\sqrt{g_{\chi}g_{q}}$ 
(where M is the mediator mass and $g_{\chi}$ and $g_q$ are the couplings to DM and quarks).

 Although this method has the great advantage to 
allow a model-independent interpretation and compare results to those from direct detection
experiments, this is not a complete theory and is valid only for sufficiently low energy:
 the transferred momentum must be low enough to be smaller than the mediator mass.

As soon as higher energies are probed this assumption does not hold any longer and EFTs breaks down. 
Models with an explicit mediator will be introduced in order to overcome this issue.



\section{Searches for dark matter production in mono-objects final states}

Most of DM searches at the LHC exploit the production of the undetected WIMP pair
 recoiling against  an object radiated by an incoming particle.
This object can be a gluon, a quark, a photon or a vector boson, and is produced 
with high transverse momentum, $p_T$ and recoils against DM leading to a quite distinctive 
signature with a high-$p_T$ object and large imbalance in transverse energy, \met.

Both ATLAS and CMS have looked for DM production in association with a mono jet~\cite{ATLASmonojet}\cite{CMSmonojet}.
Their strategy is very similar: events are required to have at least one jet with
large $p_T$ (110~GeV for CMS and 120~GeV for ATLAS) and both require large \met
Background events with genuine \met from W-decay neutrinos are suppressed by vetoing 
isolated leptons. Different strategies are employed to reject multijet events from QCD 
with non genuine \met due to jet mis-measurement. CMS allows a second jet in the event only if it is produced very close to the first one. ATLAS allows the presence of more than one jets only if they are produced fare from the missing energy.

\begin{figure}[htb]
\centering
\includegraphics[height=2.3in]{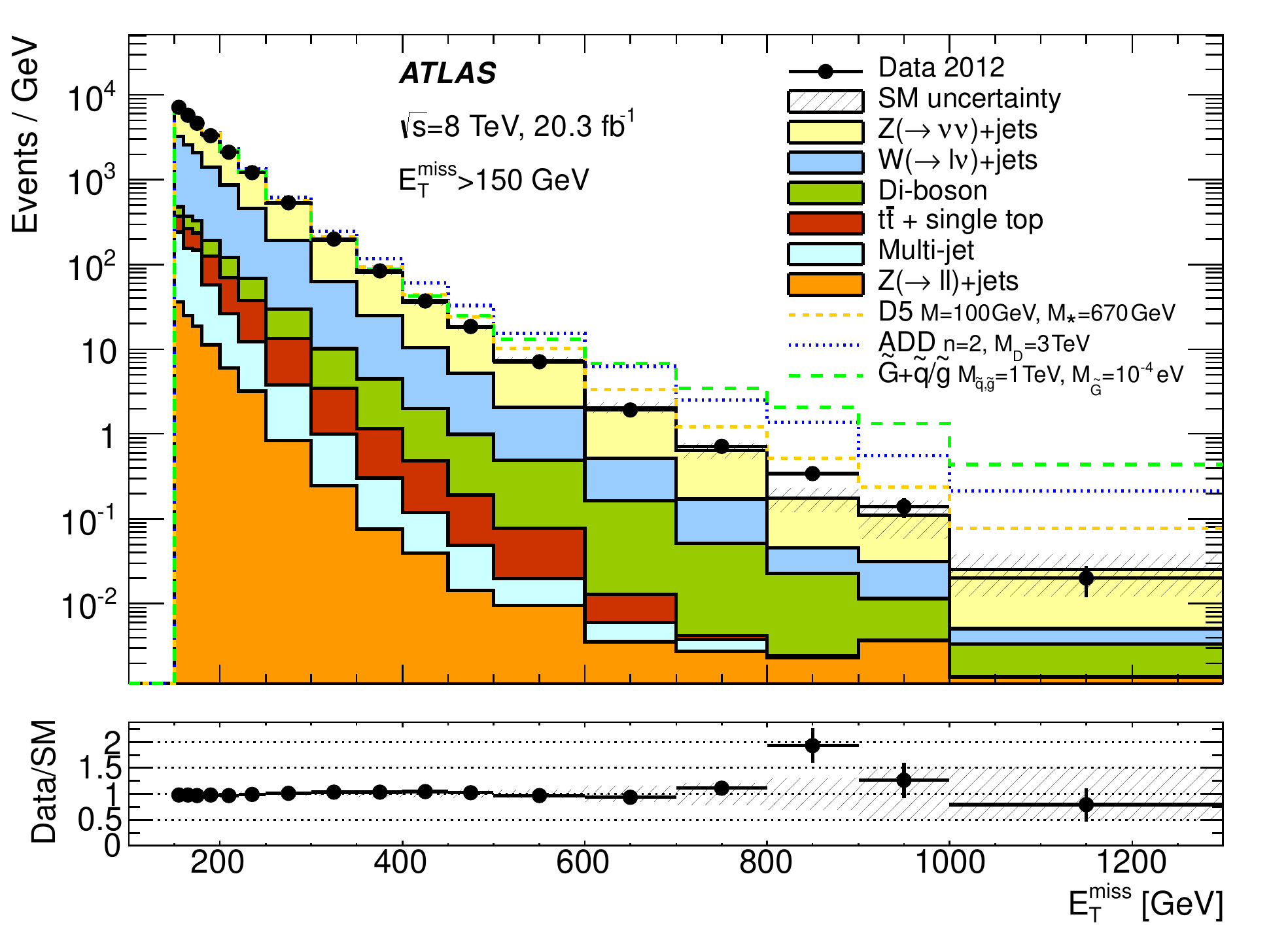}
\includegraphics[height=2.3in]{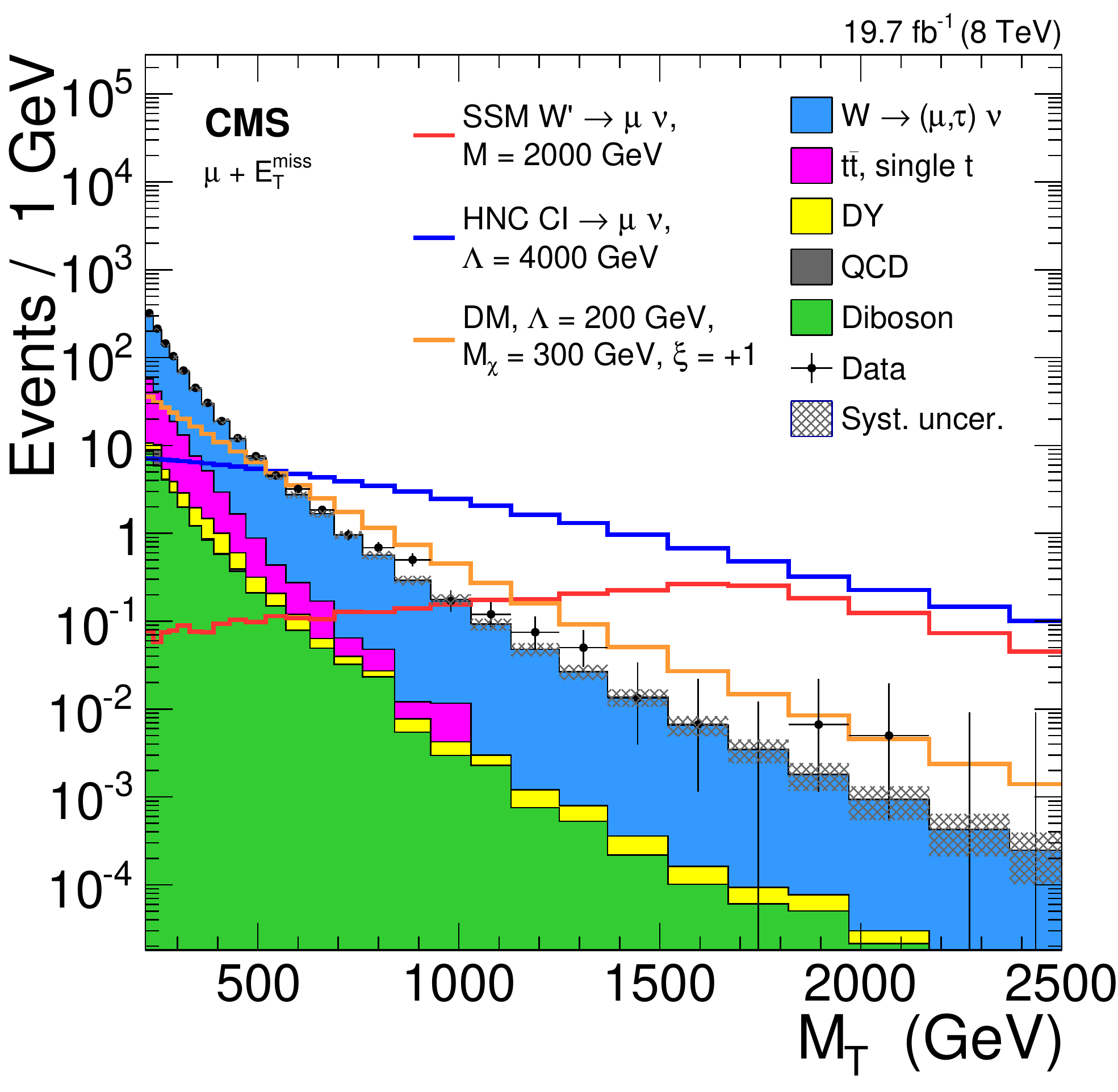}
\caption{Missing transverse energy distribution for the ATLAS monojet search (left) and transverse mass distribution for the CMS monolepton search (right) showing background expectation (filled coloured histograms) and data (black dots) in the signal region.}
\label{fig:metMonolepton}
\end{figure}

The main remaining backgrounds after election are $Z\rightarrow\nu\nu + jets$ and $W\rightarrow l\nu + jets$, 
when the lepton is not reconstructed. 
These backgrounds are estimated in the signal region by normalising the SM background
prediction to data using control regions enriched in vector bosons decaying to leptons. 
Data was found to be in good agreement with SM background expectation. In Figure~\ref{fig:metMonolepton},
the \met distribution of the most sensitive signal region is shown, comparing data with 
SM background prediction. 
with other signals from different new-physics models.
%

Although the monojet signature is the most sensitive signature for the most benchmarks
due to the large statistics in the region of interest with high \met, other DM production
processes are worth to investigate. Among them signatures with high \met and electroweak vector bosons 
have lower backgrounds with respect to monojet signature and are sensitive to different benchmark models.
Mono photon, mono-W and mono-Z searches are performed by both ATLAS and CMS.

The monophoton analysis~\cite{ATLASmonophoton}~\cite{CMSmonophoton} does not differ substantially from monojet search. 
Events with a high transverse momentum
photon ($p_T>145$~GeV) and energy imbalance (\met$>140$~GeV) are selected.
To suppress $\gamma$+jets events with \met from mis-measured jets, the selected photon is required to be back-to-back in azimuth 
with the \met ($\Delta\phi$($\gamma$, \met)$>$2).
Also lepton veto and hadron vetoes are employed to reject W+$\gamma$ and other backgrounds. 
After these requirements the Z+$\gamma$ and W+$\gamma$s constitute the dominant backgrounds in their
invisible and leptonic decays. 
No deviation from SM background expectation has been found.

Searches for DM produced in association with a weak vector boson are also performed by the two collaborations.
The monolepton search looks for DM pair produced with a W boson radiated off an incoming quark 
with the W decaying leptonically. 
Events for this signature are selected if an isolated lepton with $p_T>$~100/45~GeV (electron/muon)~\cite{CMSmonolepton}
is present. A large azimuthal opening between the lepton and the \met is imposed to discriminate signal from QCD multijets events. 
After this selection the major irreducible background is due to W$\rightarrow l \nu$ decay.
The W transverse mass, M$_T$, which has a natural endpoint at  M$_W$ if the \met originates only from W neutrino, is the main discriminating variable against this kind of background.
%
In Figure~\ref{fig:metMonolepton}(right), the M$_T$ distribution for SM background expectation and data is shown. The constructive 
interference scenario for a particular dark matter mass and energy scale is  overlaid. ATLAS performs a similar search in
this final state~\cite{ATLASmonolepton}.

\begin{figure}[htb]
\centering
\includegraphics[height=3.0in]{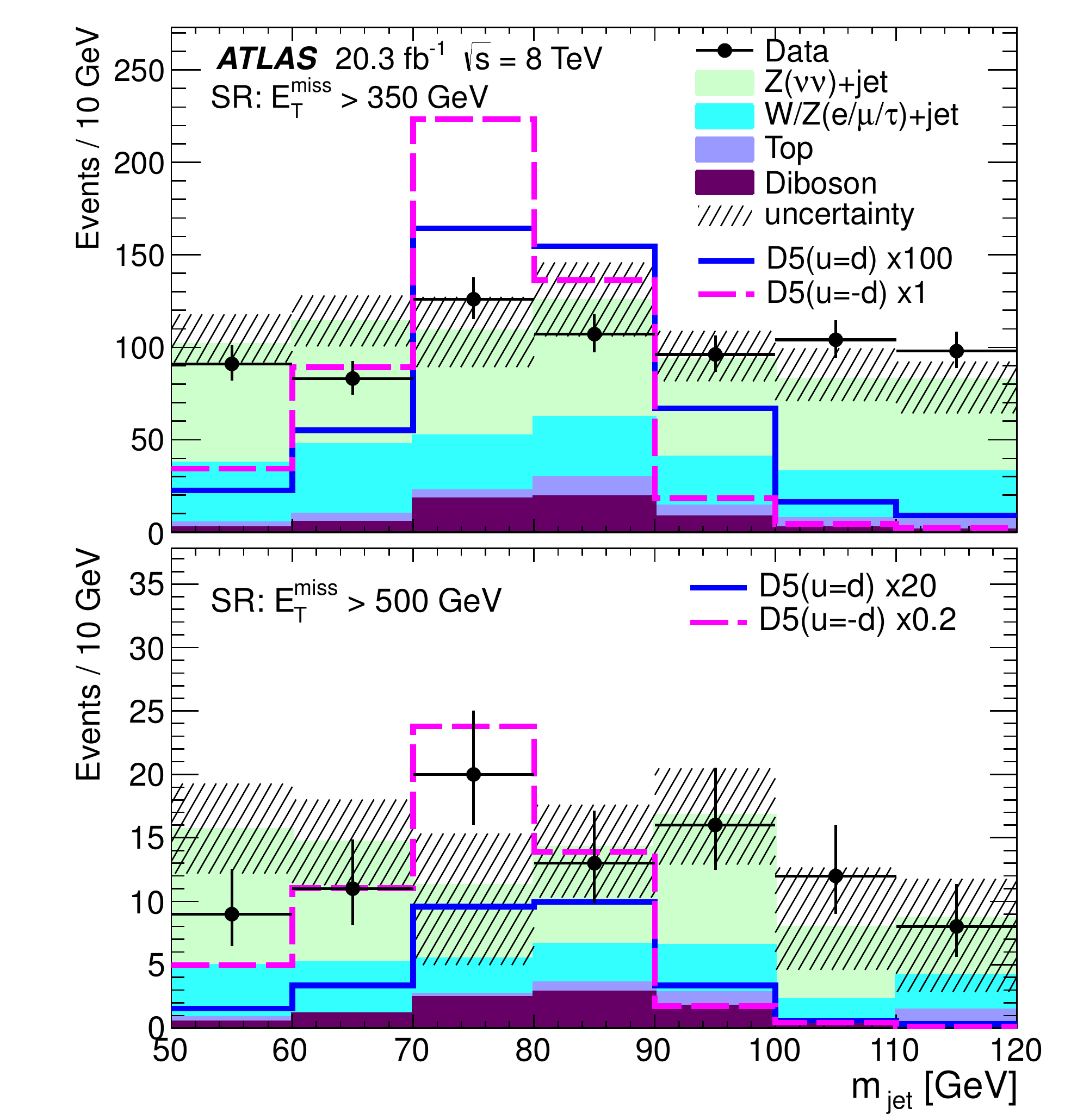}
\caption{Distribution of m$_{jet}$ for the ATLAS monoV search in the data (black dots) and for the predicted background  (filled coloured histograms) in the signal regions with \met$>350$~GeV (top) and \met$>500$~GeV (bottom). Expected signal distributions (blue and dashed magenta lines) for two DM scenarios are also shown.}
\label{fig:ATLASjetMass}
\end{figure}

Even though the lepton channel has a cleaner signature, searches in the hadronic decays of W and Z bosons take advantage 
of the large branching fraction of boson decay to a pair of jets.
ATLAS search~\cite{ATLASmonoV} look for events with a large massive jet ($fat~jet$) containing substructure consistent 
with two merged jets originated from a W or a Z boson, and with mass consistent with W/Z mass.
The fat jet mass distribution for expected backgrounds and data is represented in Figure~\ref{fig:ATLASjetMass}.
A fair agreement between data and simulation is observed in the two signal bins defined by different cuts on \met.

\subsection{Interpretation of the results}

\begin{figure}[htb]
\centering
\includegraphics[height=2.75in]{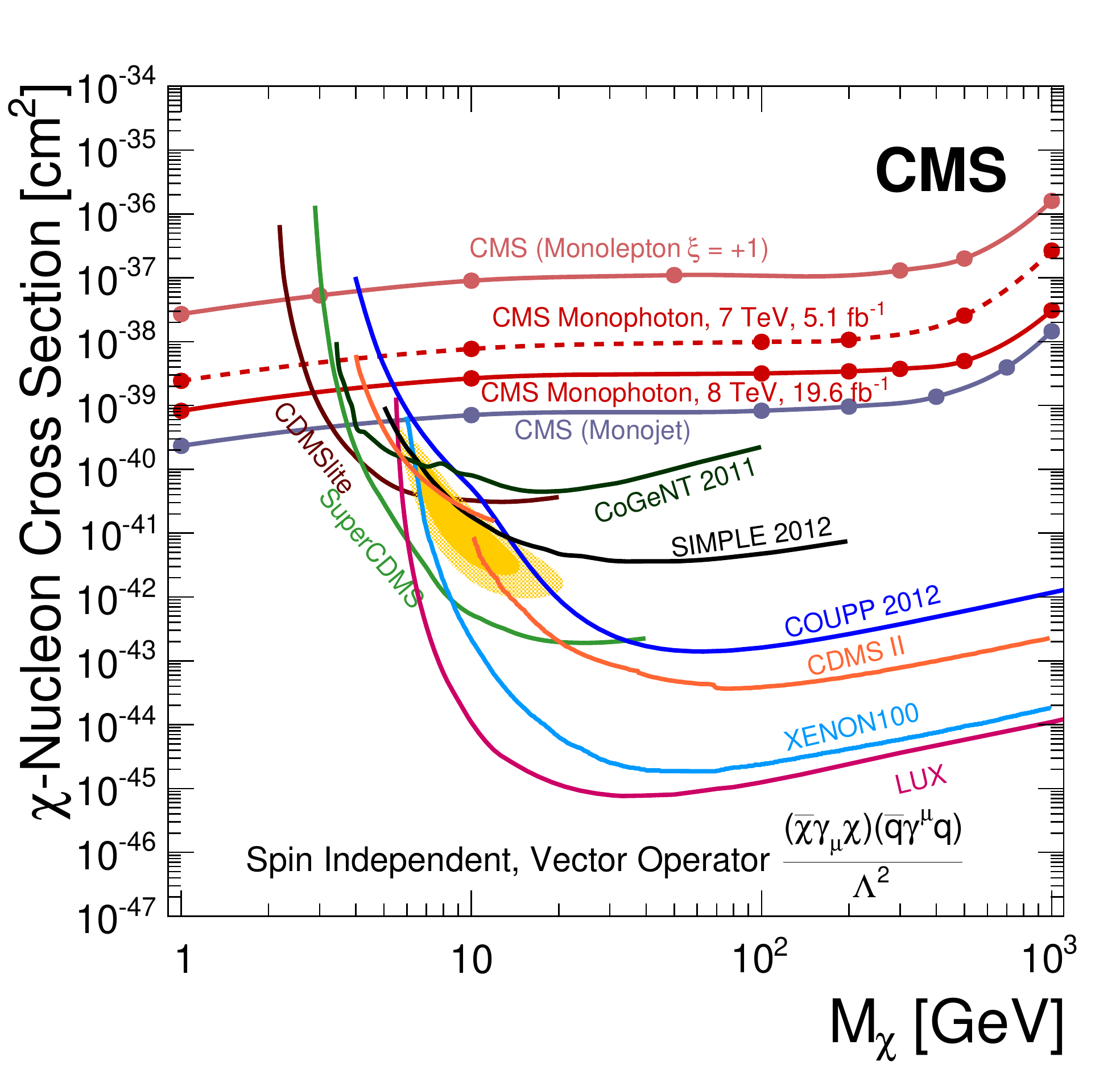}
\includegraphics[height=2.75in]{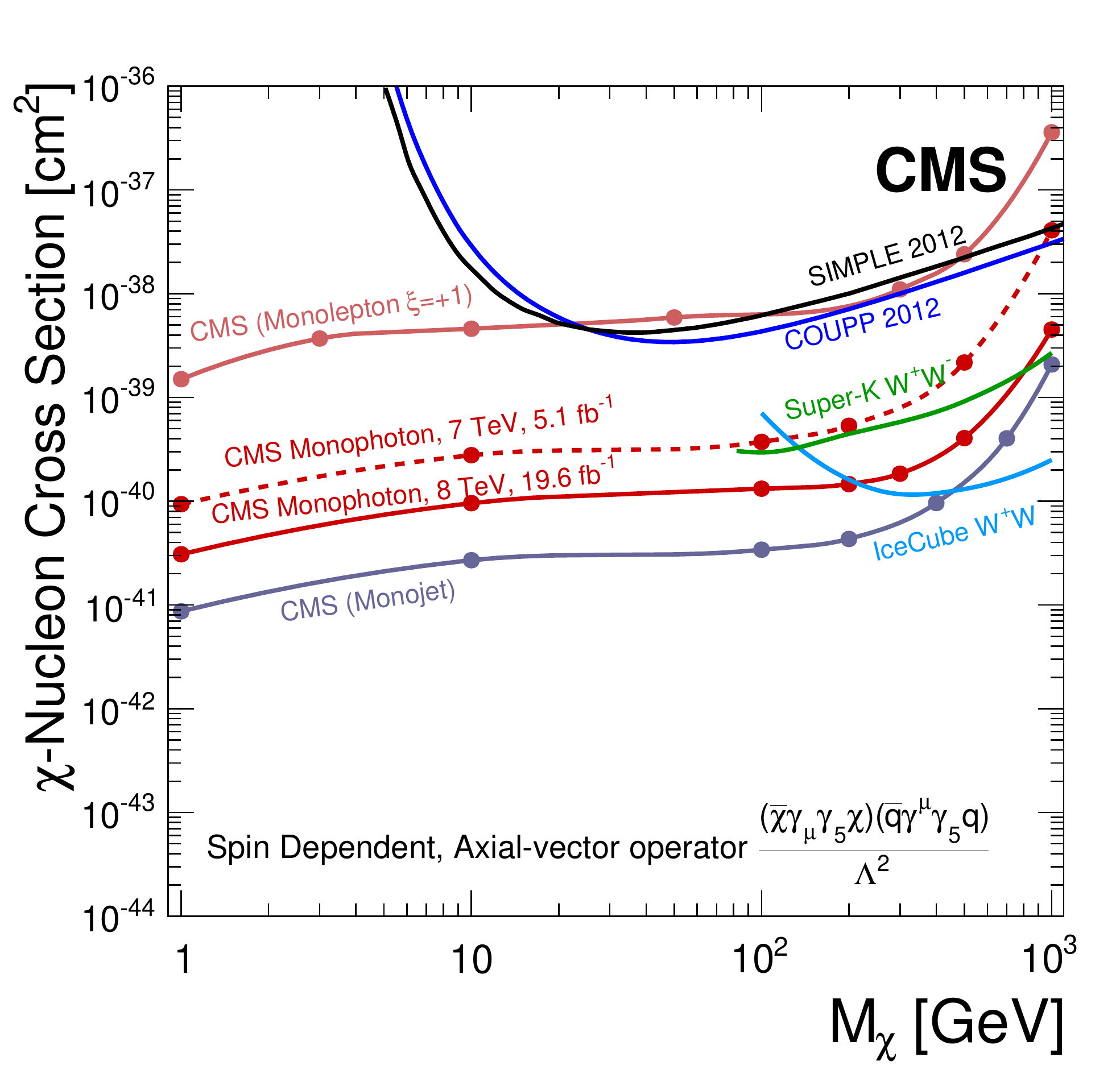}

\caption{Upper limits, at 90\% CL, on $\chi$-nucleon elastic scattering cross section as a function of m$_{\chi}$  for spin-independent (left) and spin-dependent (right) cases, compared to results from direct searches.}
\label{fig:limits}
\end{figure}

The observation in the mono-object searches presented above is found to be consistent with background-only hypothesis 
by both ATLAS and CMS. Exclusion limits are set on the DM production cross section.
These limits are then translated in lower limits on the energy scale M$^*$ of the EFT approach as a function of DM mass,
for each considered EFT interaction operator. 
A translation to the DM-nucleon elastic scattering cross section versus the dark matter particle mass plane is performed~
\cite{DMnucleonLimits} to allow a comparison of LHC results to direct detection experiments. 
The 90\% CL upper limits on the DM-nucleon scattering cross section for spin independent (vector)
and spin dependent (axial-vector) interactions is shown in Figure~\ref{fig:limits} as a function of the m$_\chi$ mass.
Comparisons are made with results from direct detection experiments showing a complementarity between collider and direct 
searches. Collider searches are more sensitive at low values of DM mass. This is due to a limitation of direct searches in 
the low mass region where the recoil signal becomes too soft to be effectively detected.
For spin-dependent interactions,  collider searches provide complementary coverage up to intermediate mass points.

\section{Searches for final states with heavy quarks}

Scalar and pseudo-scalar interactions between WIMPs and SM particles are characterised by cross sections depending on the quark mass, m$_q$. This dependence privileges the coupling of DM to heavy-flavour quarks, tops and bottoms, rather than to 
light-flavour quarks. 

Other searches for dark matter are presented to further constraints this kind of interactions: final states with missing energy  and a single or a pair of top quarks, but also with one or two bottom quarks are presented.

Many models beyond the standard model  propose the monotop production in association to an invisible state. 
Such models can be classified in two categories: resonant and non-resonant production of an invisible state.
CMS has looked at the monotop production in its hadronic decay~\cite{CMSmonotop}, while ATLAS has looked in the 
signature  with single-lepton final state~\cite{ATLASmonotop}.

The hadronic channel takes advantage of the large branching ratio of the top decay to jets.  
This search looks for events with 3 jets, out of which at least one b tagged, and with invariant mass of the three jets 
being consistent with the top mass, and large \met.
Additionally lepton veto is applied to suppress backgrounds with genuine \met ~from W decay into leptons. 
The semi-leptonic search look for events with exactly one isolated lepton and 1 b-tagged jet, together with missing energy.
The W transverse mass, M$_T$ and the azimuthal opening angle between the lepton and the b jet are constrained in order to reject background events from multijet QCD with mis-identified leptons and \met from jet mismeasurement.

In both channels data are found to be compatible with background expectation and limits on production cross section are set.


%

\begin{figure}[htb]
\centering
\includegraphics[height=2.2in]{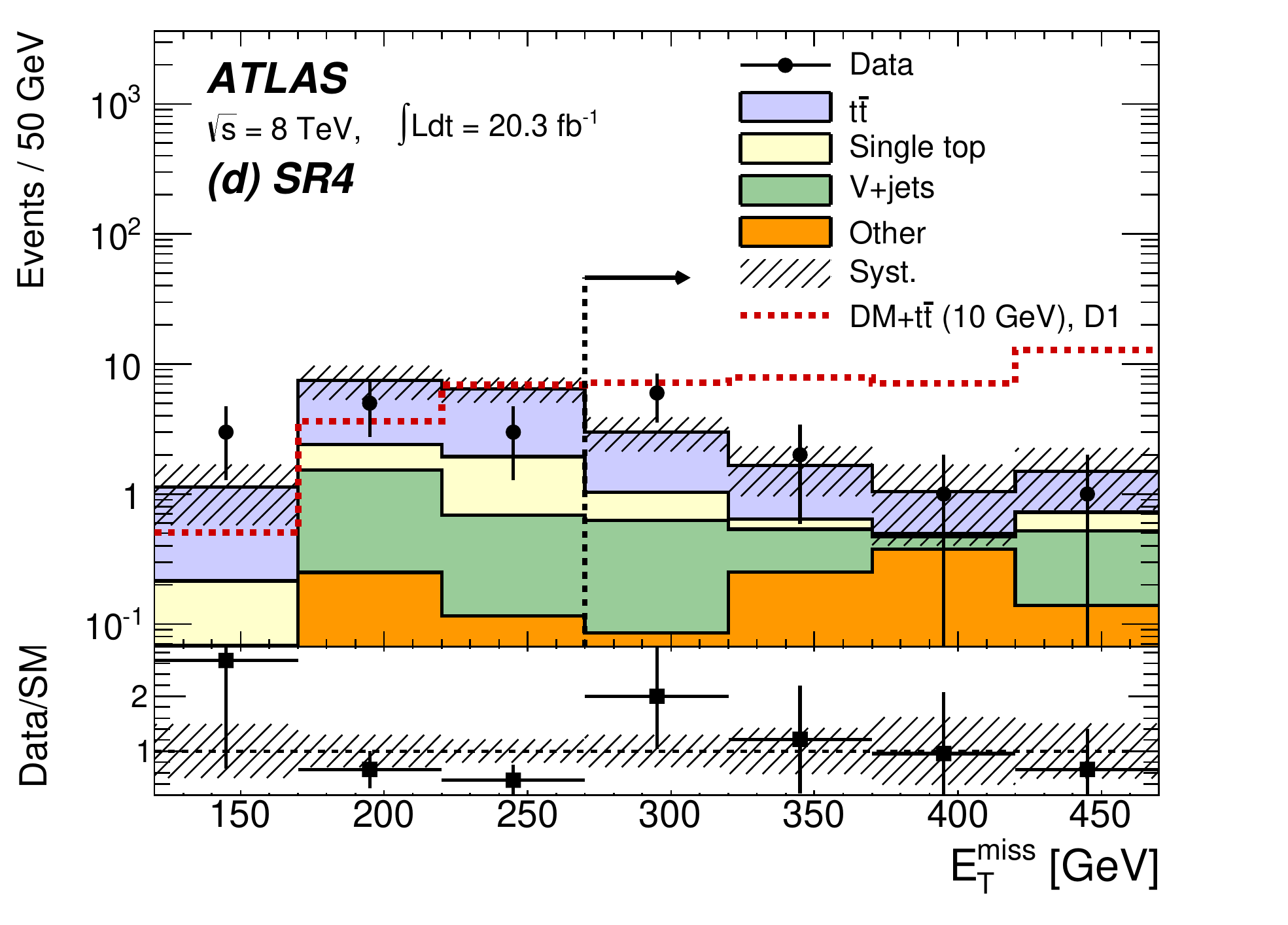}
\includegraphics[height=2.2in]{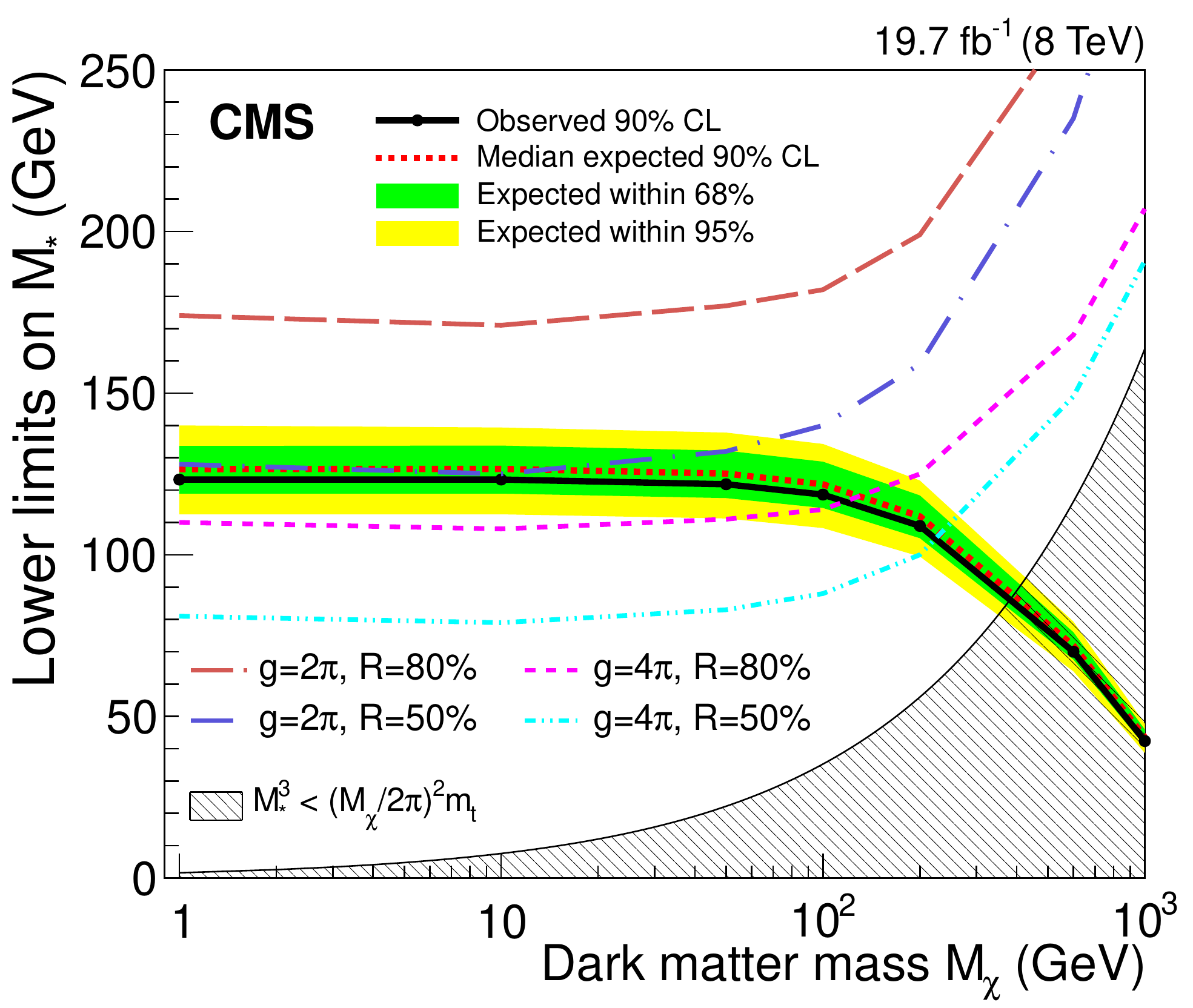}

\caption{Comparison of data and expected SM background for the \met distribution in the hadronic channel of ATLAS $t\bar{t}$+DM search, signal for a specific DM scenario is superimposed (left). Lower limits at 90\% CL on the interaction scale, M$^*$ as a function of m$_{\chi}$.}
\label{fig:ttDM}
\end{figure}

ATLAS and CMS have searched also for a DM pair produced in association with a top pair: ATLAS in the all-hadronic and single-lepton channels~\cite{ATLASttDM}, while CMS in the semileptonic and leptonic channels~\cite{CMSttDMsinglelepton}\cite{CMSttDMdilepton}. 
The most sensitive final state is the single-lepton, followed by the full hadronic and  the dilepton.
Events of this topology are identified by requiring exactly one isolated lepton in the event and at least three jets, out of which at least one b tagged.  The \met and the pair of leading jets are requested to have  large azimuthal separation, $\Delta\phi$, to suppress $t\bar{t}$ backgrounds. The main backgrounds remaining after this selection are dileptonic $t\bar{t}$ events, which are suppressed by a requirement on the M$_{T2}^W$, and a contribution from W decay, which is reduced by applying a requirement on the W transverse mass, M$_T$. 
All the backgrounds are taken from simulation. The dominant processes, $t\bar{t}$ and W+jets, are normalised to data from control regions. 
A single bin counting experiment is performed in a signal region defined by \met$>350$~GeV and lower limits on the interaction at 95\% CL are set for DM-SM scalar interaction (displayed on Figure~\ref{fig:ttDM}).
The analogous search performed by ATLAS is described in~\cite{ATLASttDM}.

Coupling of DM to bottom quarks is enhanced similarly to top quarks, thus it is worth to look for 
signatures with one or two bottom quarks produced together with DM. ATLAS and CMS uses two different approaches to select 
events of this topology. ATLAS analysis strategy~\cite{ATLASttDM} is based on the definition of signal regions based on the b-jet multiplicity, whether CMS exploits razor variables and b tagging to define signal-enriched regions~\cite{CMSbbDM}. No deviations from SM background expectations are observed and limits are set on the interaction scale assuming validity of the EFT approach. 

\section{Simplified models}


At the energies probed at the LHC the validity of the assumptions behind the EFT approach
is limited. The contact interaction validity issue together with the prospect of much higher energy 
collisions that will take place in the second Run of the LHC, leads to the need of reconsidering the benchmark models
used to interpret the DM searches in Run-I.
Considering explicitly the interaction mediator instead of integrating it out helps in overcoming the limits of 
contact interaction approach. Even though simplified models are not a complete theory, they give a more realistic 
description of the kinematics of the new process. 

\begin{figure}[htb]
\centering
\includegraphics[height=2.75in]{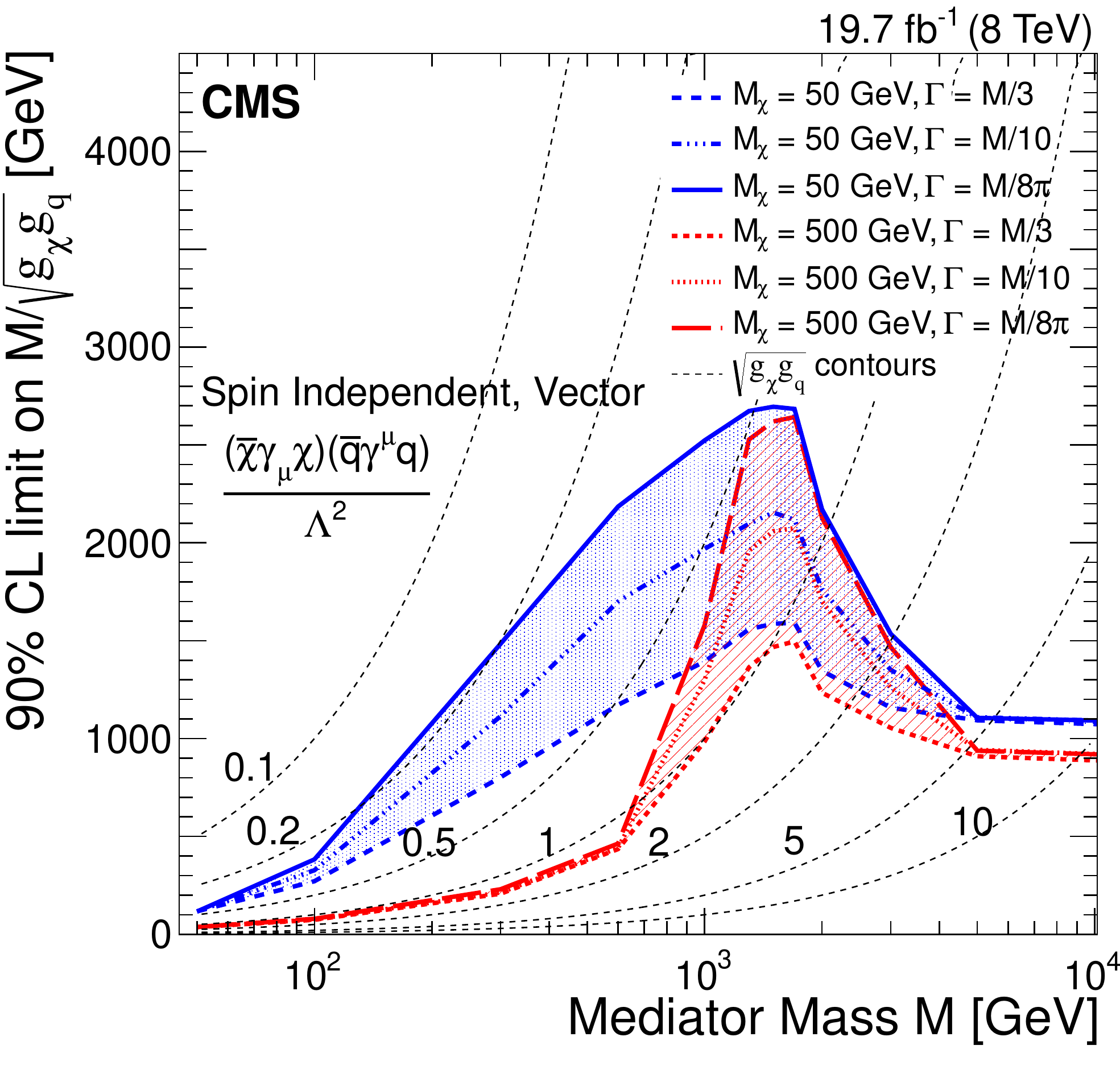}

\caption{Observed  limits, at 90\% CL, on the interaction scale M$^*$, as a function of the mediator mass, M, assuming vector interactions and a dark matter mass of 50~GeV(blue, filled) and 500~GeV (red, hatched). The width of the mediator is varied between M/3 and M/8$\pi$.  }
\label{fig:simplifiedModels}
\end{figure}

For some of Run-I searches simplified model interpretations have been considered with the mediator being a vector particle exchanged in the $s$-channel.
The limits on the interaction scale M$^*=M/\sqrt{g_{\chi}g_{q}}$ is calculated as a function of the mediator mass:
different dark matter masses and mediator width are considered. 
The results of the mediator mass scan are shown in Figure~\ref{fig:simplifiedModels} for the mono-jet search.
In the region of high mediator mass the limits from EFT and simplified model interpretations coincide.
Moving toward lower mediator masses, the mediator can be produced on-shell providing an enhancement of the
cross section due to the resonant production. The EFT approach does not take into account this enhancement and thus the limits from EFT are very conservative.
While further decreasing the mediator mass, the mediator goes of-shell again. In this regime the limit on the energy scale
from the EFT approximations are too optimistic with respect to what is found with a more realist interpretation such as 
the simplified model considered.

In view of 13~TeV LHC collisions, ATLAS and CMS collaborations have established in September 2014 a forum, the LHC DM Forum, to agree upon 
the definition of a set of simplified models for the interpretation of  results for early Run-II DM searches.
The outcome of this collaboration was summarised in the report of the Forum and made public in July~\cite{LHCDMforum}.

\section{Conclusions}

The ATLAS and CMS DM searches covered a huge range of final states during the first data-taking run of the LHC looking for signs of WIMP production.
Although observation is consistent with SM background expectation, stringent limits have been set on different benchmark models, emphasising the complementarity of collider searches and direct detection searches. 
Collider searches can powerfully constraint the low DM-mass region where the direct detection experiments suffer a lack of sensitivity. 
However the current benchmark models employed to describe the DM-SM interaction suffer of validity limitations 
in the high-energy regime. Thus a different choice will be performed for Run-II making use of simplified models which explicitly define the mediator particle, providing a more fair description of the interaction itself, and overcoming Effective Field Theory approach limitations.

\end{document}